\documentclass[aps,prx,twocolumn,superscriptaddress]{revtex4-1}
\usepackage{amsmath}
\usepackage{graphicx}
\usepackage{color}
\usepackage{amssymb}
\usepackage{subfigure}
\usepackage{float}
\usepackage{esint}
\newcommand{\figfullwidth}{16cm}
\newcommand{\figwidth}{8.6cm}

\begin{document}

\title{Optical isolation with nonlinear topological photonics}

\author{Xin Zhou}
\affiliation{Division of Physics and Applied Physics, School of Physical and Mathematical Sciences, \\
Nanyang Technological University, Singapore 637371, Singapore}

\author{You Wang}
\affiliation{Division of Physics and Applied Physics, School of Physical and Mathematical Sciences, \\
Nanyang Technological University, Singapore 637371, Singapore}

\author{Daniel Leykam}
\affiliation{Division of Physics and Applied Physics, School of Physical and Mathematical Sciences, \\
Nanyang Technological University, Singapore 637371, Singapore}

\author{Y.~D.~Chong}
\affiliation{Division of Physics and Applied Physics, School of Physical and Mathematical Sciences, \\
Nanyang Technological University, Singapore 637371, Singapore}

\affiliation{Centre for Disruptive Photonic Technologies, Nanyang Technological University, Singapore 637371, Singapore}

\email{yidong@ntu.edu.sg}

\date{\today}

\begin{abstract}
It is shown that the concept of topological phase transitions can be used to design nonlinear photonic structures exhibiting power thresholds and discontinuities in their transmittance.  This provides a novel route to devising nonlinear optical isolators.  We study three representative designs: (i) a waveguide array implementing a nonlinear 1D Su-Schrieffer-Heeger (SSH) model, (ii) a waveguide array implementing a nonlinear 2D Haldane model, and (iii) a 2D lattice of coupled-ring waveguides.  In the first two cases, we find a correspondence between the topological transition of the underlying linear lattice and the power threshold of the transmittance, and show that the transmission behavior is attributable to the emergence of a self-induced topological soliton.  In the third case, we show that the topological transition produces a discontinuity in the transmittance curve, which can be exploited to achieve sharp jumps in the power-dependent isolation ratio.
\end{abstract}

\maketitle

\section{Introduction}

Nonreciprocal light transmission plays a key role in modern optical technologies.  Optical isolators are devices that allow light to pass in one direction (e.g., along a waveguide), while blocking transmission in the other direction, thus acting as the analogues of diodes in electronic circuits.  To realize an optical isolator, the reciprocity principle of ordinary electromagnetism must be broken \cite{jalas2013}.  This can be accomplished in three distinct ways: using magneto-optic effects, temporal modulation of the electromagnetic medium, or optical nonlinearity.  Magneto-optic isolators are the most widely used in current technology, but are challenging to incorporate into on-chip optical circuits \cite{soljacic_review,el2013apl,el2015ol}.  For this reason, there has been a great deal of research into isolator designs based on spatio-temporal modulation \cite{fan2009,longhi2015cpa} and nonlinear materials \cite{soljacic_review, scalora1994, tocci1995, Gallo2001, philip2007apl, Krause2008, Poulton2010, Ramezani2010, Miroschnichenko2010, lepri2011prl, anand2013nl, lepri2013pre, xu2014prb, li2014sp, peng2014nature, chang2014nature, Fan2015, zhou2016oe, jiang2016chip, li2017ssh, Sergey2016lpr}.

This paper explores the possibility of realizing optical isolators using nonlinearity-induced topological phase transitions.  The concept of topological phases originated in the field of condensed matter physics \cite{bernevigbook}, and was introduced into photonics some years ago \cite{haldane2008prl,haldane2008pra,wang2008prl,wang2008nature,hafezi2011nphy,fang2012nphoton,liang2013,khanikaev2013nmat,rechtsman2013nature,hafezi2013,hafezi2014,lu2014nphoton}.  Researchers have demonstrated a variety of photonic structures with topologically nontrivial photonic bands, including magneto-optic photonic crystals operating at microwave frequencies \cite{haldane2008prl,haldane2008pra,wang2008prl,wang2008nature} and non-magneto-optic waveguide structures that can operate at optical frequencies \cite{hafezi2011nphy,rechtsman2013nature,hafezi2013,hafezi2014}.  Such structures possess a distinctive property: when tuned into a ``topologically nontrivial'' phase, they exhibit topological edge states that are robust against perturbations (and, in some cases, have useful properties such as unidirectionality \cite{wang2008prl}).  Although these topological edge states have mostly been studied in the linear regime, there have been several recent papers exploring how they are affected by optical nonlinearities \cite{Lumer2013,ablowitz2014,alu2016,daniel2016soliton,li2017ssh}.  It appears that photonic topological transitions---transitions from a conventional or topologically trivial phase to a topologically nontrivial phase---can be ``driven'' by nonlinearities, so that the light intensity itself determines whether the light can propagate via an edge state.  We will study how this behavior might be exploited in nonlinear optical isolators.  There have also been a number of papers seeking to implement optical isolators using magneto-optic topological photonics \cite{el2013apl,el2015ol}, but such schemes lie outside the scope of the present discussion.

We will analyze three representative photonic designs that (i) are known to exhibit topological transitions in the linear regime, and (ii) can feasibly operate in the optical frequency range, where optical isolation is a particularly pressing problem due to the absence of strong magneto-optic effects \cite{soljacic_review,el2013apl,el2015ol}.  Our goal is to obtain a conceptual understanding of the features and limitations of these novel isolation schemes; as such, we will make use of simplified models, based on the coupled-mode theory and transfer matrix frameworks, capturing just the essentials of nonlinearity and bandstructure topology.  In particular, we will not attempt to study the actual device geometries and material nonlinearities needed to achieve the nonlinear lattice parameters appearing in our models, nor to optimize our designs to maximize their performance.

The first type of structure we will study is an array of coupled optical waveguides, where light is guided (``evolves'') either forward or backward along each waveguide, and can hop between adjacent waveguides via evanescent coupling.  We begin with an exemplary waveguide array corresponding to a 1D SSH model \cite{bernevigbook}, with Kerr-like nonlinearities added to the inter-waveguide coupling strengths.  Hadad, Khanikaev, and Al\`u have shown that such a model can exhibit a self-induced topological transition \cite{alu2016}, in which the nonlinearity drives a local region of the lattice into a different topological phase, giving rise to self-trapped soliton-like edge states.  These nonlinear edge states allow a high-intensity signal injected in a edge waveguide to resist diffraction into the rest of the lattice.  We show that when such a lattice has asymmetric input and output coupling losses, it can function as an efficient optical isolator.  Light is injected into one port of an edge waveguide, evolves through a fixed distance, and leaves at the other end of the same waveguide.  With appropriately-chosen system parameters, the forward transmittance (via a self-induced topological edge state) is of order unity, while the backward transmittance (without an edge state) is suppressed by several orders of magnitude.

The isolator relies on the fact that the self-induced topological transition has a power threshold---i.e., the soliton-like edge state appears only above a certain power.  The asymmetric input and output couplings ensure that the edge state exists under forward transmission, but not backward transmission.  However, the existence of a threshold is not unique to the nonlinear SSH model.  It has previously been shown that nonlinear lattices with a conventional design can support edge solitons with a power threshold \cite{Makris2005,Suntsov2006,Suntsov2007}, distinct from bulk solitons which normally bifurcate from zero power \cite{christodoulides1988}.  In those studies \cite{Makris2005,Suntsov2006,Suntsov2007}, the edge soliton thresholds had no apparent connection to the topology of the underlying bandstructure.  In the present case, the threshold---and hence the operating power of the isolator---is set by the topological phase transition of the SSH model.

Next, we generalize these results to a structure exhibiting a 2D topological phase \cite{rechtsman2013nature}.  Unlike the 1D SSH model, the 2D lattice has edge solitons that are mobile, and can travel around defects such as corners \cite{daniel2016soliton}. In this case, to achieve strong optical isolation, the input and output must correspond to different (spatially separated) waveguides, chosen according to the nonzero velocity of the edge soliton.  Similar to the 1D case, we find that the soliton has a power threshold determined by the underlying linear model's topological phase transition---in this case, a transition from conventional insulator to Chern insulator.  This shows that the soliton-based optical isolation scheme can be extended to more complex 2D lattices.  Unlike in 1D topological phases, topologically nontrivial behavior in 2D can occur without requiring any special lattice symmetries, and is thus more robust.

The final type of nonlinear isolator that we will study is based on a periodic 2D array of coupled ring-like waveguides \cite{hafezi2011nphy,hafezi2013,hafezi2014,liang2013,liang2014,Pasek2014,Hu2015,Fei2016}.  Unlike the waveguide array case, which had a separate ``evolution'' axis $z$ (with light injected at $z=0$, and undergoing nonlinear evolution up to $z=Z$), here the system is entirely on-chip.  Light is in-coupled and out-coupled at two different positions on the lattice edge, and the steady-state solution within the nonlinear lattice is determined self-consistently.  Roughly speaking, this is like solving a nonlinear steady-state scattering problem at a fixed ``energy'', rather than a nonlinear evolution problem over a fixed ``time interval''.

We show that the coupled-ring lattice can also function as an efficient optical isolator, but with substantially different characteristics due to its steady-state nature.  The isolation is again based on a topological phase transition in the lattice bandstructure \cite{liang2013,Pasek2014}, driven by a nonlinearity-induced variation in the effective coupling between rings.  Unlike in the waveguide array, the transition manifests as a discontinuity in the nonlinear transmittance: above a critical input power, light propagation switches abruptly from very low transmittance (via bulk states) to very high transmittance (via edge states).  Thus, with varying input power, the structure is able to exhibiting a discontinuous jump from very low to very high isolation ratios.

\section{Nonlinear Coupled Waveguide Arrays: 1D SSH model}
\label{sec:array1d}

\begin{figure*}
  \centering
  \includegraphics[width=\figfullwidth]{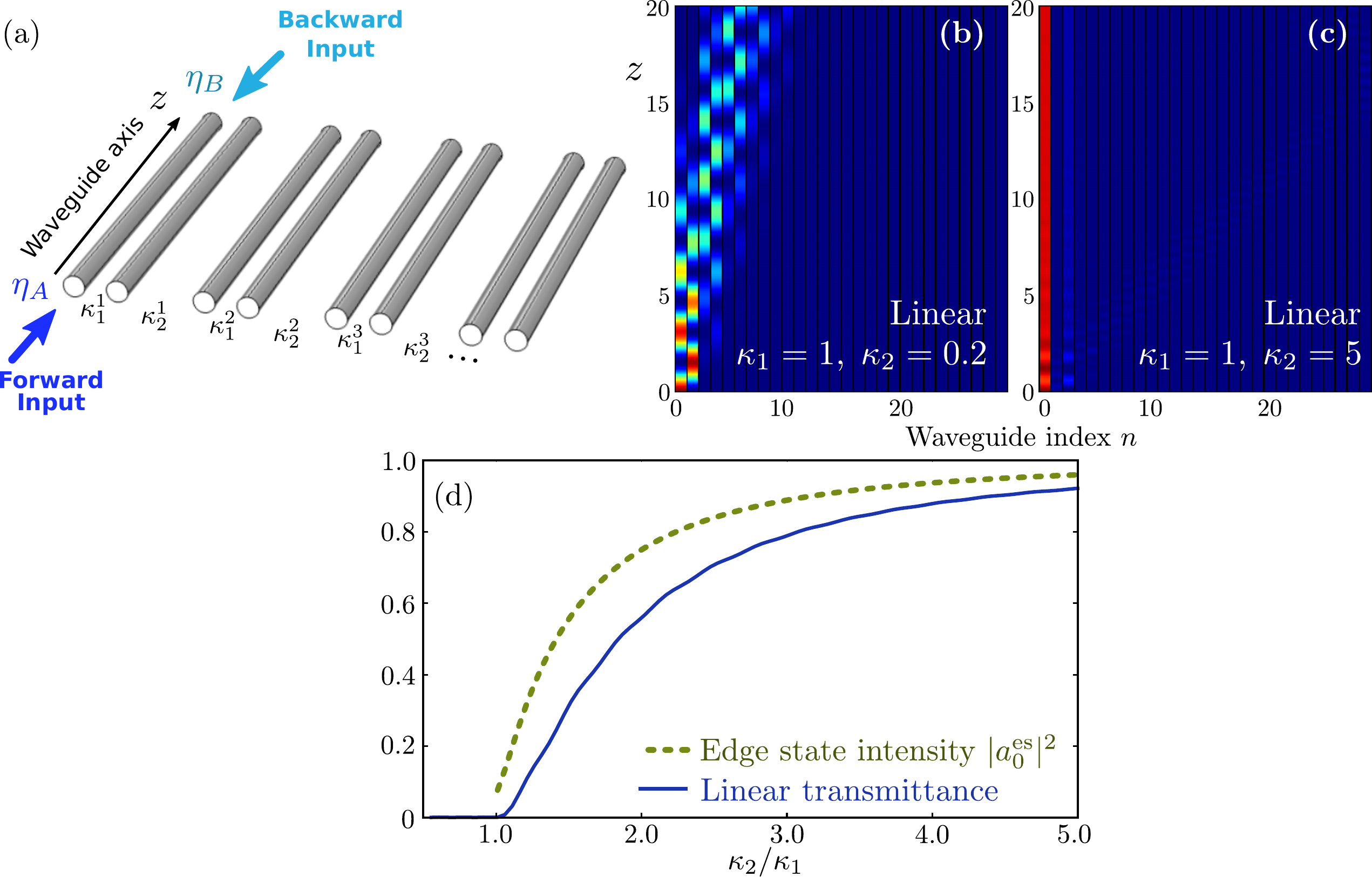}
  \caption{(a) Schematic of SSH model realized by an array of coupled waveguides.  (b,c) Mode intensity distributions in a linear SSH lattice of 29 sites, generated by an excitation on the left-most site (waveguide 1).  Results are computed numerically using the $z$-evolution operator.  (b) For a ``trivial'' lattice with $\kappa_1 = 1$ and $\kappa_2 = 0.2$, the excitation diffracts into the bulk.  (c) For a ``nontrivial'' lattice with $\kappa_1 = 1$ and $\kappa_2 = 5$, the excitation is confined to the edge.  (d) Transmittance on waveguide 1 (blue line) versus $\kappa_2/\kappa_1$.  The intensity of the edge eigenmode on waveguide 1, $|a_1|^2$, is also shown (green dashes).  In (b)--(d), the input/output couplings $\eta_A$ and $\eta_B$ are set to unity.}
\label{fig:ssh}
\end{figure*}

We begin our study with a nonlinear version of the 1D SSH model \cite{alu2016}, which is the simplest model to exhibit topological modes. As shown schematically in Fig.~\ref{fig:ssh}(a), such a model can be implemented with a 1D array of waveguides with nonlinear couplings.  Each unit cell consists of two waveguides, with identical wave-guiding characteristics.  We take the tight-binding (coupled-mode) approximation, which applies to guided modes moving in one direction along the axial direction $z$, without backscattering.  We let $a_n(z),\, b_n(z)$ denote the complex wave amplitudes in the two waveguides of unit cell $n$.  With an appropriate gauge choice, these obey  the coupled-mode equations \cite{christodoulides1988}
\begin{align}
  i\frac{da_n}{dz} &= \kappa_1^n b_n + \kappa_2^{n-1} b_{n-1}\\
  i\frac{db_n}{dz} &= \kappa_1^n a_n + \kappa_2^n a_{n+1},
\end{align}
where $\kappa_{1}^n, \kappa_{2}^n \in \mathbb{R}^+$ are intra-cell and inter-cell coupling coefficients.  In the linear regime, the $\kappa$ parameters are constants independent of $n$. The linear SSH model has been extensively investigated in photonics, including in femtosecond-laser-written waveguide arrays \cite{zeuner2015prl, weimann2016nmaterial} and plasmonic waveguide arrays \cite{cheng2015lp}.  It has a phase transition at $\kappa_1$ = $\kappa_2$; when $\kappa_1 < \kappa_2$, there is an edge state with zero eigenvalue localized to the left edge of the lattice.  We can observe this by exciting the leftmost site (waveguide 0) with input power $I = |a_0(0)|^2$, and letting the state evolve up to a fixed distance $Z$.  As shown in Fig.~\ref{fig:ssh}(b)--(c), the excitation diffracts into a superposition of bulk modes for $\kappa_2/\kappa_1 < 1$, but remains confined to the edge for for $\kappa_2/\kappa_1 > 1$.  In Fig.~\ref{fig:ssh}(d), we plot the transmittance on the edge waveguide, defined as $\mathcal{T} = |a_0(Z)|^2/I$.  This is seen to closely track the edge intensity of the exact edge eigenstate (the normalized eigenstate of the linear Hamiltonian having eigenvalue zero).

We now introduce nonlinearity into the model.  In accordance with the goals of this study, we will choose a nonlinearity that is conceptually simple and easy to model, leaving aside the question of how best to physically implement it.  The inter-cell coupling coefficient is made dependent on the local intensity, as follows \cite{alu2016}:
\begin{equation}
  \kappa_2^n(z) = \kappa_0 + \alpha \Big(|a_{n+1}(z)|^2 + |b_n(z)|^2 \Big),
  \label{ssh_nonlinearity}
\end{equation}
where $\kappa_0$ stands for static inter-cell coupling, and $\alpha$ is a Kerr-like coefficient multiplying the sum of the intensities in the two coupled sites.  We will take $\kappa_1=1.0$, $\kappa_0=0.5$, and $\alpha > 0$, i.e. the inter-cell coupling becomes stronger at higher intensities. Thus, the bandstructure is topologically trivial in the linear (zero-intensity) limit, but increasing the intensity will (roughly speaking) drive it into the nontrivial phase.  Without loss of generality, we set $\alpha = 1$ (other values are equivalent up to a rescaling of intensities).

\begin{figure}
  \centering
  \includegraphics[width=\figwidth]{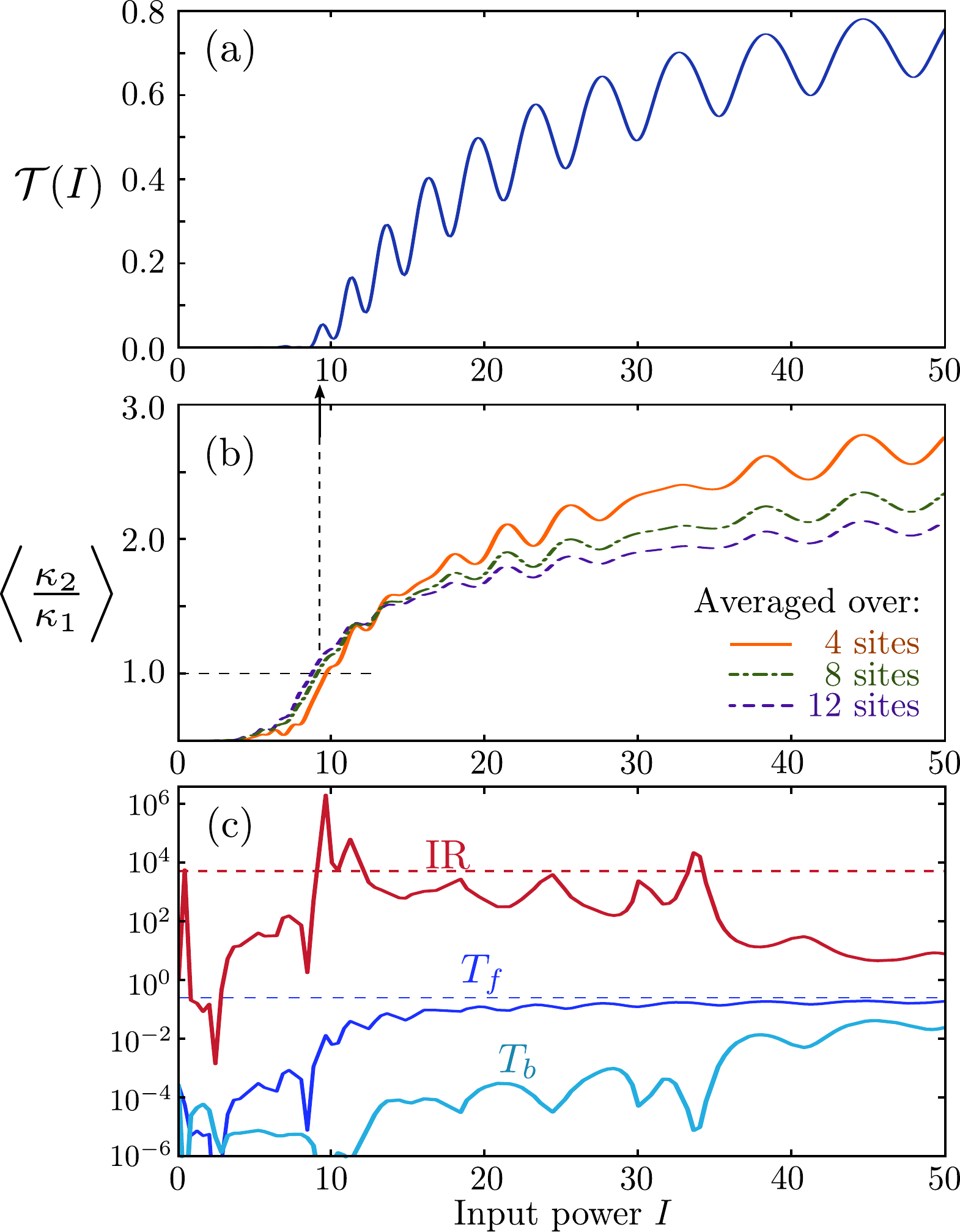}
  \caption{Behavior of the nonlinear SSH lattice as an optical isolator. (a) Nonlinear transmittance along waveguide 0, $\mathcal{T}$, versus input power $I$.  The lattice has 29 sites, with $\eta_A = 1$, $\eta_B= 0.5$, $\kappa_1=1$, $\kappa_0=0.5$, and $\alpha=1$; the output is at $Z = 20$.  Results are computed numerically using evolution operators with step size $\delta z = 10^{-3}$.  (b) Averaged values of $\kappa_2/\kappa_1$ versus input power $I$, where the nonlinear $\kappa_2$ is averaged over $18 \le z \le 22$ (i.e., around the output $Z$) and over a few sites (4, 8, or 12) closest to the left lattice edge.  The topological transition point of the linear SSH model, $\kappa_2 = \kappa_1$, is indicated by the horizontal dashes.  (c) Isolation ratio (IR), forward transmittance $T_f$, and backward transmittance $T_b$, versus input power $I$.  Horizontal dashes show the approximate upper bounds $T_{f,b}^{\mathrm{max}} = \eta_A^2 \eta_B^2$ and $\mathrm{IR}^{\mathrm{max}} \approx 1/T(0)$.}
\label{fig:isolation}
\end{figure}

The nonlinearity allows for the possibility of a self-induced topological transition \cite{alu2016}.  Suppose we prepare an initial state by exciting just the left edge waveguide with input power $I = |a_0(0)|^2$.  The light undergoes nonlinear evolution for distance $Z$, and we compute the transmittance $\mathcal{T}(I) = |a_0(Z)|^2/I$.  The results are shown in Fig.~\ref{fig:isolation}(a).  For small $I$, $\mathcal{T}(I)$ is close to zero (similar to the linear case, the light mostly diffracts into the lattice bulk); but for $I \gtrsim 9$, $\mathcal{T}(I)$ abruptly (but continuously) increases towards unity.

This abrupt change in the nonlinear transmittance is related to a self-induced topological transition. To see this, in Fig.~\ref{fig:isolation}(b) we plot the averaged values of $\kappa_2/\kappa_1$ as a function of $I$ (since $\kappa_2$ varies between sites and also with $z$, the averages shown here are taken over different numbers of unit cells near the left lattice edge, and over a fixed range of $z$).  We find that the increase in $\mathcal{T}(I)$ starts to occur when $\kappa_2/\kappa_1 \approx 1$, which is precisely the topological transition point of the SSH model.  Note that correspondence is apparent even when we average $\kappa_2$ over 12 sites, a relatively far distance from the edge.

The regime of high nonlinear transmittance is due to the self-induced edge soliton described by Hadad \textit{et al.}~\cite{alu2016}.  This is a nonlinear mode that inherits some properties of the linear SSH model's edge state, such as leaving the $b_n$ sites unexcited.  Its onset also closely matches the topological transition of the SSH model. However, it differs in other ways: its eigenvalue is not pinned to zero (as the SSH model's ``particle-hole'' symmetry is broken by the nonlinearity), and the intensity goes to a small constant in the bulk rather than decaying exponentially with distance from the edge \cite{alu2016}. 

A nonlinear photonic structure with strongly intensity-dependent transmittance can serve as the basis for an optical isolator \cite{philip2007apl, anand2013nl, peng2014nature, chang2014nature, zhou2016oe, jiang2016chip}.  To accomplish this, we introduce couplings $\eta_A$ and $\eta_B$, as shown schematically in Fig.~\ref{fig:ssh}(a).  In forward-transmission mode, light is coupled into port $A$ on an edge waveguide, is guided in the $+z$ direction, and is out-coupled at port $B$ of the same edge waveguide.  In backward-transmission mode, light enters at $B$, is guided in the $-z$ direction, and is out-coupled at port $A$.  (Note that this is distinct from ``asymmetric light transmission'' schemes, such as those studied in Ref.~\onlinecite{longhi2015cpa}, where light propagates either left-to-right or right-to-left along the lattice, while being guided in a single axial direction $+z$; those schemes have no direct bearing on the problem of optical isolation, as they do not swap physical input and output ports.)  In forward-transmission mode, the input light has intensity $I$, and the intensity coupled into the edge waveguide is $\eta_A^2 I$; the intensity at the end of the waveguide is $\mathcal{T}(\eta_A^2 I) \, \eta_B^2 \, I$, where $\mathcal{T}$ is the nonlinear transmittance of the lattice itself.  The overall forward transmittance is thus
\begin{equation}
  T_f = \eta_A^2 \, \eta_B^2 \; \mathcal{T}(\eta_A^2 I).
  \label{eq:Tf}
\end{equation}
Similarly, the backward transmittance (in the $-z$ direction) is
\begin{equation}
  T_b = \eta_A^2 \, \eta_B^2 \; \mathcal{T}(\eta_B^2 I).
  \label{eq:Tb}
\end{equation}
Note that $T_f = T_b$ in the linear regime $I = 0$, in accordance with the recprocity principle.  The isolation ratio is defined by
\begin{equation}
  \mathrm{IR} \;\equiv\; T_f/T_b\; =\;
  \frac{\mathcal{T}(\eta_AI)}{\mathcal{T}(\eta_BI)}.
\end{equation}

Fig.~\ref{fig:isolation}(c) shows numerical results for the forward and backward transmission, and the isolation ratio, for $\eta_A = 1$ and $\eta_B = 0.5$.  We can understand these results qualitatively by using Eqs.~(\ref{eq:Tf})--(\ref{eq:Tb}) and the features of the nonlinear waveguide transmittance $\mathcal{T}(I)$.  First, note that the upper bound for the transmittance (in either direction) is $T_{f,b}^{\mathrm{max}} \le \eta_A^2\eta_B^2$, which occurs when the light is transmitted predominantly along the edge, with losses only at the input and output ports.  As $I$ is increased from zero, the transmittances increase exponentially from the very low value of $\eta_A^2\eta_B^2\mathcal{T}(0)$, before saturating near the upper bound.  Since $\eta_A \ne \eta_B$, the initial increase of $T_f$ and $T_b$ occur with different exponential factors, and as a result the IR increases exponentially with the input power $I$.  It reaches a maximum of $\mathrm{IR}^{\mathrm{max}} \approx 1/\mathcal{T}(0)$, where $\mathcal{T}(0)$ is the waveguide transmittance in the linear regime.  This corresponds to the case where $T_f$ has saturated but not $T_b$.  The results in Fig.~\ref{fig:isolation}(c) show good agreement with these approximate bounds.

We can compare these results to optical isolation schemes based on ``non-topological'' edge solitons, such as solitons induced by on-site Kerr nonlinearity~\cite{Makris2005,Suntsov2006,Suntsov2007}.  Such solitons also exhibit a power threshold proportional to the coupling $\kappa$. Optimizing the isolation ratio in a device of fixed length thus requires a trade-off between minimizing the linear transmittance (larger $\kappa$), or minimizing the threshold power (smaller $\kappa$). In contrast, in the nonlinear SSH model one can reduce the linear transmittance by increasing $\kappa_{0,1}$ without substantially effecting the threshold power, determined by $\kappa_1-\kappa_0$. Moreover, the staggered profile of the topological edge soliton enables further optimization of the performance, for example by incorporating lossy elements onto the unexcited $b_n$ sites to further reduce the backward transmission~\cite{el2015ol}.

\section{Nonlinear Coupled Waveguide Arrays: 2D Haldane model}
\label{sec:array2d}

\begin{figure*}
  \centering
  \includegraphics[width=\figfullwidth]{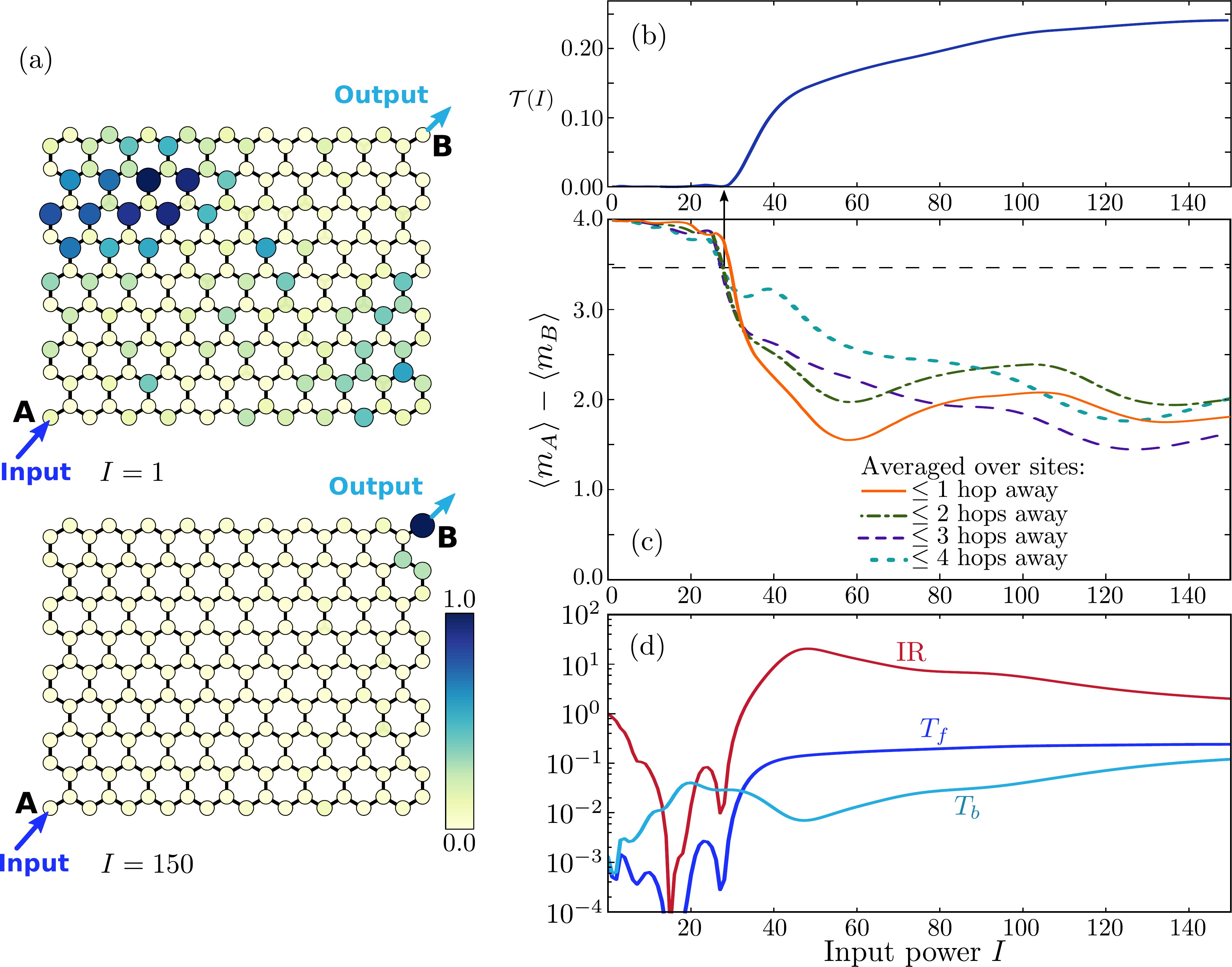}
  \caption{(a) Intensities after evolution through a nonlinear Haldane lattice, with parameters $t_1 = 1$, $t_2 = 1/3$, $\phi = \pi/2$, and $m_0 = 2$.  A corner site (A) is initially excited with intensity $I$, and colors indicate site intensities $|a_\mu|^2$ and $|b_\nu|^2$ (with arbitrary normalization) after propagation through $Z = 19.2$. Two cases are shown: $I = 1$ and $I = 150$.  The input/output couplings $\eta_A$ and $\eta_B$ are set to unity.  (b) Nonlinear transmittance $\mathcal{T}(I)$ from one corner (A) to the opposite corner (B), versus input power $I$.  (c) Mean values of $\langle m_A\rangle - \langle m_B\rangle$, versus $I$.  The averages are taken over sites up to $n$ nearest-neighbor hops from site B, for $n \in \{1,2,3,4\}$.  (d) Isolation ratio (IR), forward transmittance $T_f$, and backward transmittance $T_b$, versus $I$. }
\label{fig:haldane}
\end{figure*}

The phenomenon of optical isolation aided by a topological transition can also be observed in 2D lattices.  There are two important difference between 1D and 2D.  Firstly, topological edge states in 2D can exhibit unidirectional propagation along the edge, so we choose the input and output ports to be different waveguides.  Secondly, whereas topological protection in the SSH model requires a specific sublattice symmetry, topological protection in 2D generally does not; hence, 2D lattices could provide robust isolation under a wider range of fabrication imperfections or nonlinearities.

In the linear regime, it has previously been shown that a 2D optical waveguide array can be made to act as a 2D topological insulator with broken ``time-reversal'' symmetry (a Chern insulator) by adding a helical twist to the waveguides \cite{rechtsman2013nature}.  A variant design has been shown to support tunable topological phase transitions \cite{leykam2016a}.  Here, we leave implementation details to one side, and focus instead on the Haldane model \cite{Haldane1988}, the simplest and most well-known 2D model with a topological phase transition between a conventional insulator phase and a Chern insulator phase \cite{Haldane1988}.  This tight-binding model describes a 2D honeycomb lattice with broken time-reversal symmetry; the honeycomb lattice is divided into two sub-lattices, $A$ and $B$, with on-site mass terms $m_A = m_0$ and $m_B = -m_0$.  The other model parameters are the nearest-neighbor hopping $t_1 \in \mathbb{R}$, the next-nearest-neighbor hopping amplitude $t_2 \in \mathbb{R}$, and an Aharanov-Bohm phase $\phi \in [-\pi,\pi]$ which determines the magnetic flux penetrating sub-regions of each unit cell \cite{Haldane1988}.  The bandstructure is in a Chern insulator phase when
\begin{equation}
  \left| m_0/t_2 \right| < 3\sqrt{3} \left|\sin\phi\right|,
  \label{haldane_phases}
\end{equation}
with Chern numbers $C = \mathrm{sgn}(\phi)$.  The other phase is a conventional insulator with Chern number $C = 0$.

We now introduce a nonlinearity designed to drive the system through a topological transition.  We make the on-site mass terms on the $A$ and $B$ sublattices nonlinear, depending on the local intensity, as follows:
\begin{align}
  \begin{aligned}
  m_A^\mu = \frac{m_0}{1+|a_\mu|^2}\\
  m_B^\nu = \frac{-m_0}{1+|b_\nu|^2},
  \end{aligned}
  \label{haldane_nonlinearity}
\end{align}
where $m_0 \in \mathbb{R}$ is the Haldane model's mass parameter in the linear limit; $\mu,\nu$ are site indices on the $A$ and $B$ sublattices respectively; and $a_\mu, b_\nu \in \mathbb{C}$ denote the optical wave amplitude (wavefunction) at those sites.  Similar saturable nonlinearities have previously been studied in the context of non-topological photonic lattices with mobile discrete solitions \cite{Melvin2006}.  We will take $\phi = \pi/2$, $t_2 = 1/3$, and $m_0 = 2$, so that according to Eq.~(\ref{haldane_phases}) the linear system is in the trivial insulator phase.  With increasing intensity, the saturable nonlinearity decreases the on-site mass parameters, ``driving'' the system towards the Chern insulator phase.

The resulting nonlinear propagation is shown in Fig.~\ref{fig:haldane}(a).  Input light is injected onto a single site at a corner of the lattice, with intensity $I = |a_0|^2$, and undergoes nonlinear evolution over a fixed distance $Z$; as discussed in Section~\ref{sec:array2d}, the $z$ axis plays the role of time.  For small $I$, the light diffracts into the bulk of the lattice, in accordance with the fact that the bandstructure of the linear system is topologically trivial.  For large $I$, the light remains tightly confined to the edge, and propagates in one direction along the edge, including around corners, in accordance with the existence of a unidirectional edge state in the Chern insulator phase.

The nonlinear transmittance $\mathcal{T}(I)$ is plotted in Fig.~\ref{fig:haldane}(b), and it exhibits a power threshold similar to what we observed in the nonlinear SSH model.  Here, $\mathcal{T}(I)$ is defined as the transmittance from the input site (A) at $z = 0$, to the output site (B) on the opposite lattice corner at $z = Z$.  Again, we can demonstrate a close correspondence between the power threshold and the topological transition of the linear lattice.  In Fig.~\ref{fig:haldane}(c), we plot $\langle m_A\rangle - \langle m_B\rangle$ versus $I$, where $\langle m_A\rangle$ and $\langle m_B\rangle$ are the nonlinear on-site mass terms averaged over sites closest to the output waveguide at $z = Z$.  Based on Eq.~(\ref{haldane_phases}), the linear lattice exhibits a topological transition at $m_A - m_B = \left|6\sqrt{3}t_2 \sin\phi\right| = 2\sqrt{3}$; in Fig.~\ref{fig:haldane}(c), we indeed observe that the nonlinear transmittance's threshold power occurs as $\langle m_A\rangle - \langle m_B\rangle$ drops below this value.  In Fig.~\ref{fig:haldane}(d), we plot the isolation ratio and forward and backward transmittances.  The behavior is similar to the nonlinear SSH results in Fig.~\ref{fig:isolation}(c).  One difference is that $T_f$ and $T_b$ are dissimilar even though the input/output couplings are symmetric (here we set $\eta_A = \eta_B = 1$); this is because the inputs and outputs are on different sites, so the lattice itself provides the asymmetry.

Optical isolation based on traveling edge solitons has important qualitative differences compared to schemes based on immobile solitons, such as the nonlinear SSH model discussed in Section~\ref{sec:array1d}.  The device length $Z$ and/or the choice of output waveguide must be matched to the edge soliton velocity, so as to ensure a high forward transmittance. This kind of traveling discrete soliton is not easily achievable with ``conventional'' nonlinear lattice designs not tied to a topological transition; in those cases, traveling solitons require excitation of several waveguides~\cite{Melvin2006}, and discrete solitons are typically immobile and/or suffer from strong radiative losses~\cite{Morandotti1999}.

\section{Nonlinear coupled ring lattices}

\begin{figure}
  \centering
  \includegraphics[width=\figwidth]{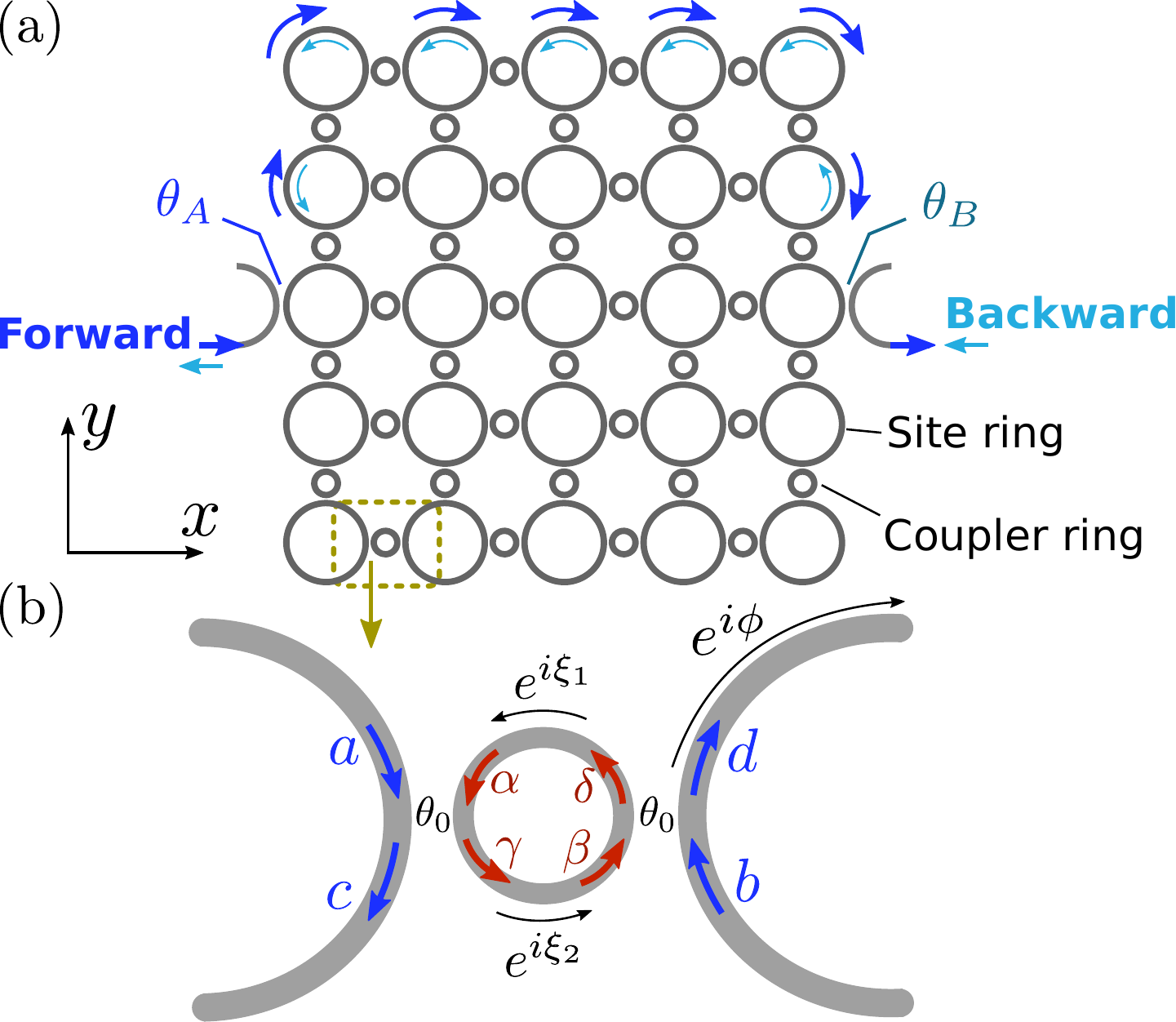}
  \caption{(a) Schematic of a lattice of coupled rings, consisting of ``site rings'' (large circles) arranged in a square lattice, separated by auxiliary ``coupler rings'' (small circles).  Light is coupled in/out of the lattice on the left and right, with coupling parameters $\theta_A$ and $\theta_B$.  (b) Close-up of an inter-site coupling, showing the definitions of the complex wave amplitudes $\{a, b, c, d\}$ (in the site rings) and $\{\alpha, \beta, \gamma, \delta\}$ (in the coupler ring). }
  \label{fig:ringresonators}
\end{figure}

We now turn our attentions to a quite different type of photonic structure: a lattice of nonlinear coupled rings.  The structure is shown schematically in Fig.~\ref{fig:ringresonators}(a), and consists of ring-shaped waveguides arranged in a 2D square lattice, with adjacent ``site rings'' connected by auxiliary ``coupler rings''.  The structure is assumed to be engineered so that there is negligible back-scattering at the coupling regions where  neighboring waveguides approach one another; in other words, the circulation of light in the site rings---clockwise or anti-clockwise---is preserved under inter-site hopping \cite{hafezi2011nphy,hafezi2013,hafezi2014}.  

Suppose the entire lattice, including the auxiliary rings, is periodic.  In this case, the lattice's bandstructure is known to exhibit a topological transition \cite{liang2013}: as we increase the effective coupling between site rings, the bandstructure goes from a conventional phase to a topologically nontrivial phase.  In the latter, there exists (for each circulation) a family of topological edge states that move unidirectionally along the lattice edge \cite{liang2013,liang2014,Pasek2014,Hu2015,Fei2016}.

For the waveguide arrays discussed in Section~\ref{sec:array1d}--\ref{sec:array2d}, the ``forward'' and ``backward'' modes of the optical isolator corresponded to $+z$ and $-z$ propagation.  By contrast, the present coupled-ring lattice structure is ``on-chip'', i.e.~purely 2D.  In forward-transmission mode, light is coupled into one ring, propagates through the lattice in a given circulation direction (say, clockwise), and is subsequently out-coupled.  In backward-transmission mode, the input and output ports are switched, and hence the in-lattice propagation takes place via the \textit{opposite} circulation (anti-clockwise).  It is important to note that this switches the directionality of the topological edge states.  For instance, in Fig.~\ref{fig:ringresonators}(a) we show a right-moving edge state on the upper edge, with clockwise circulation; the reciprocal partner is a left-moving edge state on the upper edge, with anti-clockwise circulation.

The transmission of light through the lattice can be calculated using transfer matrices \cite{hafezi2011nphy,liang2013}.  Fig.~\ref{fig:ringresonators}(b) shows a coupler ring joining two site rings; the complex wave amplitudes labeled in this figure are related by
\begin{equation}
  \begin{bmatrix} c \\ \gamma \end{bmatrix}
  = S_c(\theta_0) \begin{bmatrix} a\\ \alpha \end{bmatrix}, \;\;
  \begin{bmatrix} d\\ \delta \end{bmatrix}
  = S_c(\theta_0) \begin{bmatrix} b \\ \beta \end{bmatrix},
\end{equation}
where
\begin{equation}
  S_c(\theta_0) = \begin{bmatrix}
    \sin\theta_0&i\cos\theta_0\\i\cos\theta_0&\sin\theta_0
  \end{bmatrix}
  \label{Sc}
\end{equation}
is a unitary $2\times2$ scattering matrix describing evanescent coupling with energy conservation and negligible back-scattering, as well as $180$-degree rotational symmetry \cite{hafezi2011nphy, liang2013, liang2014}.  The strength of the evanescent coupling is described by $\theta_0$.  Moreover, wave amplitudes acquire a phase $\xi_{1,2}$ on traversing each arm of the coupler ring:
\begin{equation}
  \alpha = e^{i\xi_1}\, \delta, \; \beta = e^{i\xi_2} \, \gamma.
\end{equation}
The effective coupling between adjacent site rings can be determined \cite{liang2013,liang2014,Pasek2014} from the parameters $\theta_0$ and $\xi_{1,2}$ (which depend, in turn, on the waveguide geometry and operating frequency).  We will assume that the couplings in the $x$ and $y$ direction are identical.

\begin{figure}
  \centering
  \includegraphics[width=\figwidth]{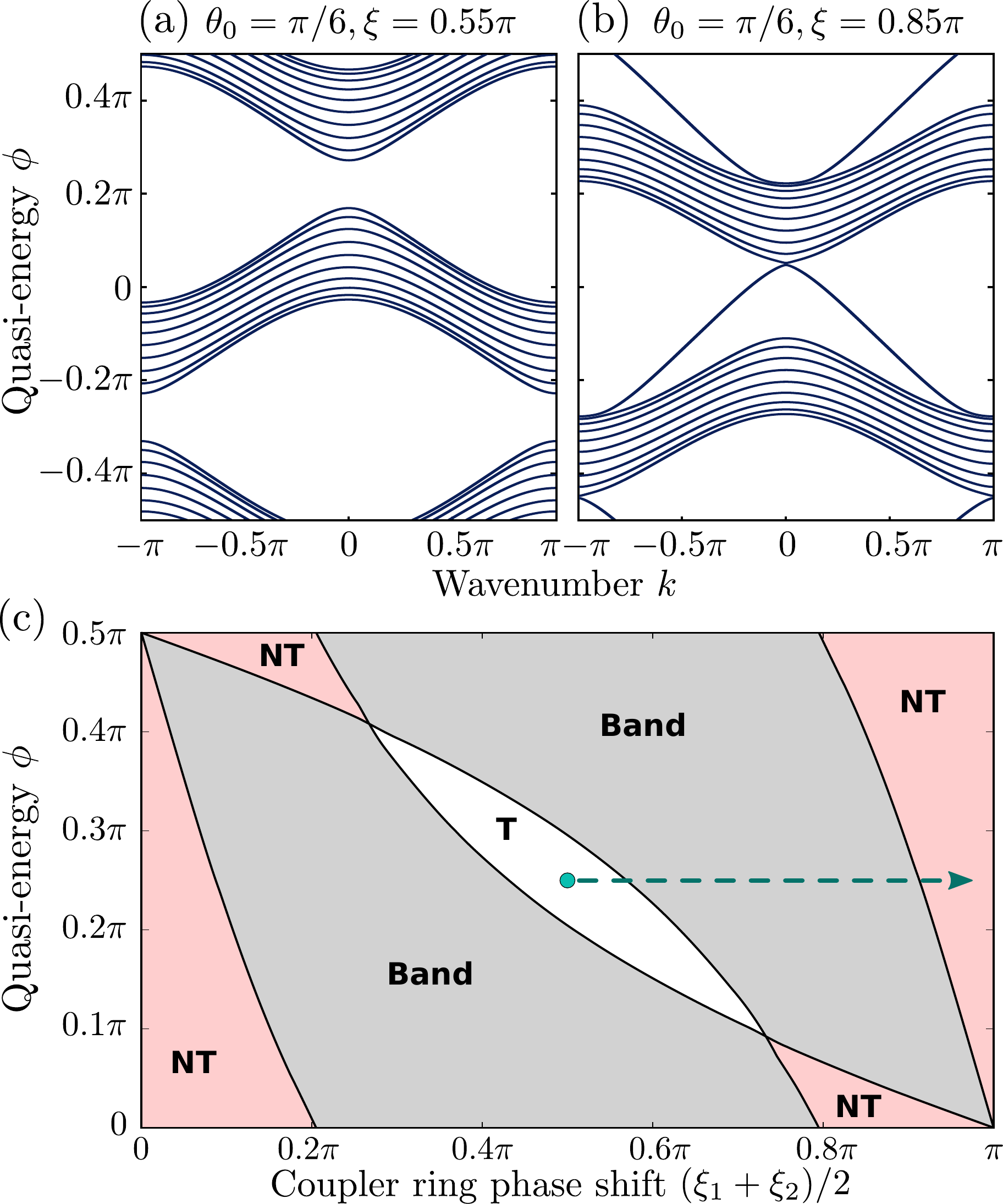}
  \caption{(a)--(b) Band diagrams for a semi-infinite lattice of coupled rings, with width 10 unit cells in the $y$ direction.  For the coupler ring parameters, we choose $\theta_0 = \pi/6$ and two different choices of the phase shift $\xi$; the resulting bandstructures are (a) topologically trivial for $\xi_1 = \xi_2 = \xi = 0.55\pi$, and (b) topologically nontrivial for $\xi = 0.85\pi$.  (c) Numerically-obtained phase diagram of the system for $\theta_0 = \pi/6$.  Grey regions show the ungapped parts of the bandstructure, white regions show trivial (T) bandgaps, and pink regions show nontrivial (NT) bandgaps.  For fixed $\phi = \pi/4$, the system can move from a trivial to a nontrivial gap by varying $\xi_{1,2}$ as indicated by the arrow.  The gaps depend only on $\xi_1$ and $\xi_2$ through the combination $\xi_1 + \xi_2$.}
  \label{fig:Bands}
\end{figure}

The bandstructure is defined in terms of the phase shift $\phi$ over each quarter of a site ring, as indicated in Fig.~\ref{fig:ringresonators}(b).  This plays the role of a ``quasi-energy'' \cite{liang2013,Pasek2014}, whose value is fixed by the waveguide geometry and operating frequency; we can regard $\phi$ as being analogous to the chemical potential in a band insulator.  (Note that the other model parameters, $\theta_0$ and $\xi_{1,2}$, will also simultaneously depend on the operating frequency; in designing a real device, these model parameters must be mapped onto physical quantities including the frequency and geometrical parameters.)  In Fig.~\ref{fig:Bands}(a)--(b), we plot band diagrams of $\phi$ versus $k$ for a semi-infinite lattice with transverse width of 10 unit cells; here, $k$ is the usual Bloch wavenumber, defined as the phase shift produced by translating one unit cell along the strip \cite{liang2013}.  We fix $\theta_0=\pi/6$, and show results for two choices of $\xi = \xi_1 = \xi_2$.  Fig.~\ref{fig:Bands}(a) shows a conventional bandstructure, while Fig.~\ref{fig:Bands}(b) shows a nontrivial bandstructure with gaps spanned by topological edge states.  Evidently, we can switch between these two distinct cases by tuning only $\xi$, with all other model parameters kept constant.  Fig.~\ref{fig:Bands}(c) shows a phase diagram indicating the parameter choices for observing trivial and nontrivial gaps.

\begin{figure}
  \centering
  \includegraphics[width=\figwidth]{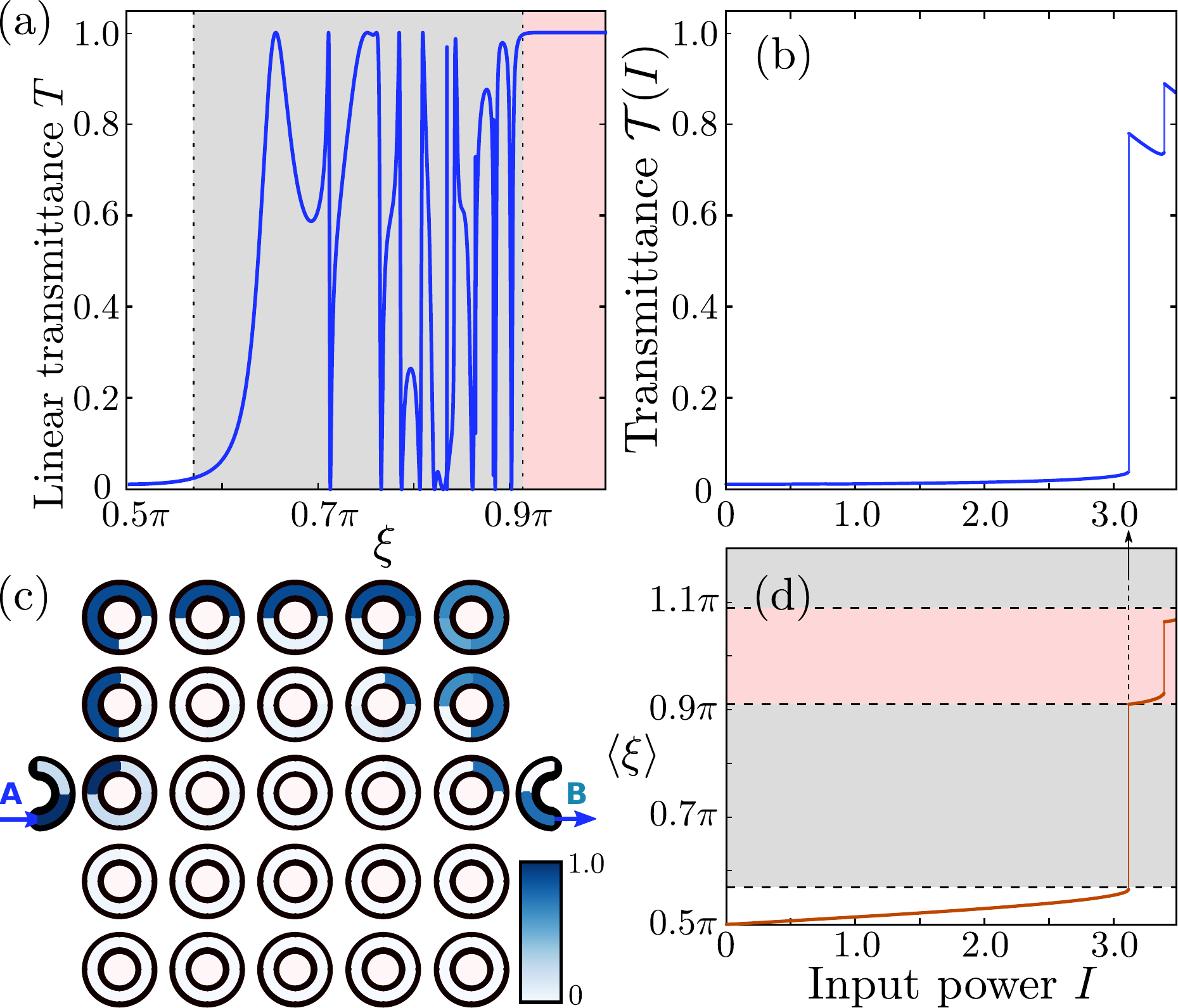}
  \caption{(a) Transmittance $T$ through a linear lattice with phase shift $\xi$ in all arms of all coupler rings.  The input/output couplings are taken to be perfect (the coupling matrices have the form of Eq.~(\ref{Sc}) with $\theta_A = \theta_B = \pi/2$), and the lattice parameters are $\phi = \pi/4$, and $\theta_0 = \pi/6$.  For $\xi \approx \pi/2$, $\phi$ lies in a trivial gap, and $T$ is close to zero; with increasing $\Delta \xi$, $\phi$ enters a band (grey region) and $T$ exhibits bulk transmission resonances; for still larger $\xi$, $\phi$ enters a nontrivial gap (pink region) and $T$ approaches unity.  Vertical dashes indicate the boundaries between the band and gap regions shown in Fig.~\ref{fig:Bands}(c).  (b) Transmittance through the nonlinear lattice versus input power $I$, using nonlinearity parameters $\xi_0 = \pi/2$ and $\kappa = 1$, with all other parameters kept the same as in (a).  This and subsequent results are obtained by solving the nonlinear transfer matrix relations self-consistently via a numerical nonlinear solver.  (c) Normalized field intensity distribution at $I = 3.21$, just above the large discontinuity in (b).  For clarity, the coupler rings are omitted and only the site ring intensities are shown. (d) Mean phase shifts $\langle\xi\rangle = \langle(\xi_1+\xi_2)/2\rangle$ versus $I$, with averages taken over both arms of the eight coupler rings on the upper lattice edge [see Fig.~\ref{fig:ringresonators}(a)].  Horizontal dashes indicate the boundaries from Fig.~\ref{fig:Bands}(c).}
  \label{fig:Trans_I_psi}
\end{figure}

Fig.~\ref{fig:Trans_I_psi}(a) shows the transmittance across the linear lattice for fixed $\phi = \pi/4$ and varying coupler ring phase shifts $\xi = \xi_1 = \xi_2$.  The transmittance is close to zero for $\xi \approx \pi/2$, when the system is in a trivial gap, and approaches unity when $\xi$ increases and the system enters a topologically nontrivial gap.  For intermediate $\xi$, the system lies in a band, and the transmittance exhibits numerous resonances.

We now introduce a nonlinearity designed to drive the system through the topological transition.  Let the phase shift parameter on each arm of a coupler ring be intensity-dependent:
\begin{equation}
  \xi = \xi_0 + \kappa \mathcal{I},
  \label{eq:kerr_coupler}
\end{equation}
where $\xi_0$ is the phase shift in the linear limit, $\kappa$ is a Kerr coefficient, and $\mathcal{I}$ is the local intensity in the arm of the coupler ring, defined as $\mathcal{I} = |\psi|^2$ where $\psi$ is the local complex amplitude.  Physically, this may be accomplished by fabricating the coupler rings out of a nonlinear material (note that the intensity within the coupler rings will be strongly enhanced if they are close to resonance with the operating frequency \cite{liang2014}).  We choose $\phi = \pi/4$, $\theta_0 = \pi/6$ (both assumed to be intensity-independent), $\xi_0 = \pi/2$, and $\kappa = 1$.  Referring to the phase diagram in Fig.~\ref{fig:Bands}(c), we see that in the linear limit the bands are topologically trivial, with $\phi$ lying within a trivial gap.  Increasing $\mathcal{I}$, and hence $\xi$, drives the system (locally) into a nontrivial gap.  In interpreting this phase diagram, note that although the nonlinearity changes $\xi$ independently in the two arms of a coupler ring, the band gaps of the linear system depend only on the \textit{sum} of $\xi$ in the two arms.

Fig.~\ref{fig:Trans_I_psi}(b) plots the transmittance across the lattice, $\mathcal{T}(I)$, versus input power $I$.  Here, we assume the input and output couplings to be perfect (the coupling matrices have the form of Eq.~(\ref{Sc}), with coupling angle $\pi/2$).  These results are obtained by using a standard numerical nonlinear solver to find self-consistent solutions to the entire set of transfer matrix relations within the lattice, including the nonlinear phase shifts described by Eq.~(\ref{eq:kerr_coupler}).  From Fig.~\ref{fig:Trans_I_psi}(b), we observe that the transmittance in the linear limit, $\mathcal{T}(0)$, is negligible, in accordance with the fact that the system is in a trivial gap.  Moreover, with increasing $I$, $\mathcal{T}(I)$ initially remains low, but at a critical intensity it jumps discontinuously to $\mathcal{T} \approx 0.8$.  Above this discontinuity, transmission takes place along the upper lattice edge, as shown in Fig.~\ref{fig:Trans_I_psi}(c).

In Fig.~\ref{fig:Trans_I_psi}(d), we plot $\langle\xi\rangle$ versus $I$, averaging over the coupler rings on the upper lattice edge.  This shows that the large discontinuity in $\mathcal{T}(I)$ found in Fig.~\ref{fig:Trans_I_psi}(b) occurs at values of $\langle\xi\rangle$ corresponding to the boundary between the trivial gap and the band, and between the band and the nontrivial gap.  By comparison with the linear system's phase diagram from Fig.~\ref{fig:Bands}(c), it appears that the nonlinear lattice is ``jumping'' past the in-band regime.

Fig.~\ref{fig:Trans_I_psi}(b) and (d) also shows a ``secondary'' discontinuity, where $\langle\xi\rangle$ jumps to a larger value while remaining in the topological gap region.  This seems to be triggered when the nonlinearity causes the value of $\xi$ in a few individual coupler rings to exceed the upper boundary of the topological gap region, entering into another in-band region at $\xi \sim 1.1\pi$.  This destabilizes the solution, and the system compensates by jumping to a different field distribution, with larger $\langle\xi\rangle$, that keeps all coupling rings in the topological gap region.  On both sides of this secondary discontinuity, the field distribution remains confined to the upper lattice edge.

\begin{figure}
  \centering
  \includegraphics[width=0.8\linewidth]{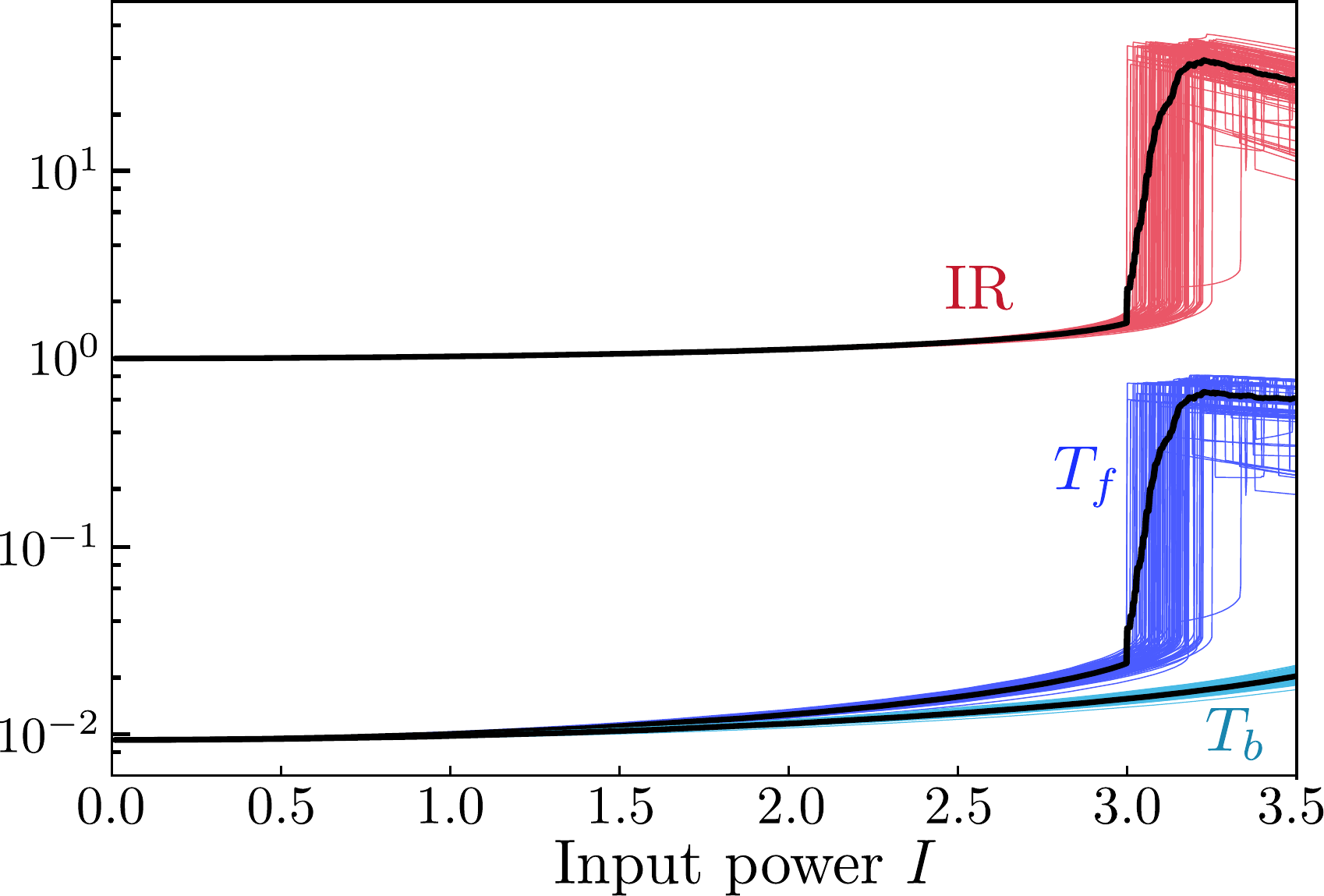}
  \caption{Isolation ratio (IR), forward transmittance $T_f$, and backward transmittance $T_b$, versus input power $I$, in disordered nonlinear coupled ring lattices.  Thin lines show results for 100 different disorder realizations, and thick black lines show the mean values.  The disorder consists of shifts in individual coupler ring phase shifts $\xi_j$, normally distributed with mean 0 and standard deviation $0.01\pi$.  Input/output loss is applied at port B by multiplying amplitudes by $e^{0.1}$ (with no loss at port A); all other parameters are the same as in Fig.~\ref{fig:Trans_I_psi}.}
  \label{fig:IRP}
\end{figure}

Unlike the nonlinear Haldane model studied in Section~\ref{sec:array2d}, there is no ``time-of-flight'' limitation on the propagation of light from the input to the output port, due to the steady-state nature of the model.  As such, optical isolation in the nonlinear coupled-ring lattice can exhibit topological protection against lattice defects.  To test this idea, we apply an input/output asymmetry by adding a small amount of loss (coupling factor of $e^{-0.1}$) to one of the ports (B).  We then compute the self-consistent forward transmittance (port A to B), backward transmittance (B to A), and isolation ratio.  Disorder is introduced by adding normally distributed shifts to each $\xi_i$ parameter (with standard deviation $0.01\pi$).  The results are shown in Fig.~\ref{fig:IRP}.  For a range of input powers near $I \gtrsim 3$, there is a sharp jump in the mean forward transmittance $\langle T_f\rangle$, and correspondingly a jump in the mean isolation ratio.  These jumps come from ensemble averages of transmittance discontinuities shifted by the disorder; note that the secondary discontinuities, being much smaller in magnitude, are ``smeared out'' and are thus not visible in the plotted mean values.

\section{Discussion and conclusions}

In this paper, we have studied how optical isolation can be accomplished in three different models of topological photonics.  A few basic ingredients are common to all three models.  Firstly, the model must be nonlinear, so as to break optical reciprocity \cite{jalas2013}.  Secondly, the structure must contain an asymmetry (e.g., asymmetric input/output couplings) that distinguishes between ``forward'' and ``backward'' directions.  The third ingredient is the use of a topological phase transition to associate forward transmission and backward transmission with different topological phases, whose physical properties are qualitatively different from each other.  This last design principle is reminiscent of recently-proposed optical isolation schemes based on parity/time-reversal (PT) symmetric structures, which rely on a non-Hermitian transition (between ``PT symmetric'' and ``PT broken'' phases) rather than a topological phase transition \cite{peng2014nature,chang2014nature,zhou2016oe, Sergey2016lpr}.

In the presence of nonlinearity, topological phase transitions manifest as ``self-induced'' topological solitons: local regions of the lattice where the optical field self-sustains its own edge state-like behavior, even when the lattice is topologically trivial in the zero-intensity limit.  Such solitons were previously discovered in the 1D nonlinear SSH model \cite{alu2016}, as well as 2D continuum ``Floquet topological insulator'' models \cite{Lumer2013,daniel2016soliton}.  Here, we have shown that the nonlinear SSH soliton is useful for optical isolation, and moreover that topological solitons also occur in two new 2D models---nonlinear versions of the Haldane model \cite{haldane2008prl} and the coupled-ring lattice \cite{hafezi2011nphy,hafezi2013,hafezi2014,liang2013,Pasek2014}.  The soliton in the coupled-ring lattice model is notable for being a static (steady-state) solution, whereas the solitons in the nonlinear Haldane model and other previously-studied 2D lattices \cite{Lumer2013,daniel2016soliton} are dynamical.  As we have seen, this gives rise to the distinctive feature: a discontinuity (not just a threshold or kink) in the power-dependent optical transmittance.

To realize a self-induced topological soliton, nonlinearity must be applied to a model in a non-arbitrary way: as exemplified by Eqs.~(\ref{ssh_nonlinearity}), (\ref{haldane_nonlinearity}), and (\ref{eq:kerr_coupler}), the nonlinearity needs to act on parameters that drive the system towards a topological phase transition.  Interestingly, homogeneous nonlinearities such as local Kerr effects, which are the most commonly-studied nonlinearities in lattice models \cite{Makris2005,Suntsov2006,Suntsov2007,christodoulides1988}, may \textit{not} be suited to inducing topological phase transitions.  Roughly speaking, such uniform nonlinearities play the role of altering the scalar potential, which is an inefficient way to induce topological band inversions.  In real experiments, nonlinearities may be inhomogeneous and present in both on-site terms and ``off-diagonal'' (inter-site coupling) terms~\cite{nonlinear_lattices}.  Our results demonstrate that the latter, though frequently ignored, can lead to soliton behaviors that are both distinctive and useful.  It should also be noted that our study has omitted the temporal effects of optical nonlinearity, such as frequency generation and pulse dispersion; these may be important in certain materials, or for ultrashort pulses~\cite{boyd_book}. 

In future studies, it would be interesting to introduce additional features to the models that could further improve their performance as optical isolators.  For instance, loss can be selectively added to the lattice to suppress the transmittance from bulk modes and/or diffusive non-topological edge modes, while leaving the topological edge modes relatively unaffected.  It would also be interesting to make a comparison with various non-topological nonlinear isolator designs that are fine-tuned to achieve high isolation ratios and/or high transmittance~\cite{lepri2011prl,anand2013nl,lepri2013pre,xu2014prb}; one might be able to show that isolation schemes based on topological solitons can achieve similarly high performance, while being less sensitive to random defects due to the intrinsic robustness of the topological edge modes.

Dynamical effects in the coupled-ring lattice are another avenue for further study. For example, the transmittance jumps in Fig.~\ref{fig:Trans_I_psi}(b) may form hysteresis loops under an additional slow modulation of the input intensity.  Such designs may also be useful for limiting unwanted dynamical reciprocity~\cite{fan2009}, as the nonlinear isolation is provided by topological edge states localized in both frequency and space. Small amplitude signals with sufficiently large frequency detuning from the input could propagate via the qualitatively different bulk modes, and might thus be efficiently filtered or suppressed. 

We are grateful to M.~Hafezi and S.~Mittal for their helpful comments and suggestions.  This research was supported by the Singapore MOE Academic Research Fund Tier 2 (grant MOE2015-T2-2-008) and the Singapore MOE Academic Research Fund Tier 3 grant MOE2011-T3-1-005.


\begin{thebibliography}{99}
\bibitem{jalas2013}
  D.~Jalas \textit{et al.}, \textit{What is--and what is not--an optical isolator}, Nat.~Phot.~\textbf{7}, 8 2013.

\bibitem{soljacic_review} M.~Solja\u{c}i\'{c} and J.~D.~Joannopoulos,
  \textit{Enhancement of nonlinear effects using photonic crystals},
  Nat.~Mat.~\textbf{3}, 211 (2004).

\bibitem{el2013apl}
  R.~El-Ganainy, M.~Levy, A.~Eisfeld, and D.~N.~Christodoulides,
  \textit{On-chip non-reciprocal optical devices based on quantum inspired photonic lattices}, Appl.~Phy.~Lett.~\textbf{103}, 161105 (2013).

\bibitem{el2015ol}
  R.~El-Ganainy and M.~Levy, \textit{Optical isolation in topological-edge-state photonic arrays}, Opt.~Lett.~\textbf{40}, 5275 (2015).

\bibitem{fan2009}
Yu, Zongfu, and Shanhui Fan, \textit{Complete optical isolation created by indirect interband photonic transitions}, Nature photonics~\textbf{3}, 91-94 (2015).

\bibitem{longhi2015cpa}
  S.~Longhi, \textit{Non-reciprocal transmission in photonic lattices based on unidirectional coherent perfect absorption}, Opt.~Lett.~\textbf{40}, 1278 (2015).

\bibitem{scalora1994} M.~Scalora, J.~P.~Dowling, C.~M.~Bowden, and M.~J.~Bloemer, \textit{The photonic band edge optical diode}, J.~Appl.~Phys.~\textbf{76}, 2023 (1994).

\bibitem{tocci1995} M.~D.~Tocci, M.~J.~Bloemer, M.~Scalora, J.~P.~Dowling, and C.~M.~Bowden, \textit{Thin‐film nonlinear optical diode}, Appl.~Phys.~Lett.~\textbf{66}, 2324 (1995).

\bibitem{Gallo2001} K.~Gallo, G.~Assanto, K.~R.~Parameswaran, and M.~M.~Fejer,
  \textit{All optical diode in a periodically poled lithium niobate waveguide},
  Appl.~Phys.~Lett.~\textbf{79}, 314 (2001).

\bibitem{philip2007apl} R.~Philip, M.~Anija, C.~S.~Yelleswarapu, and D.~V.~G.~L.~N.~Rao, \textit{Passive all-optical diode using asymmetric nonlinear absorption}, Appl.~Phys.~Lett.~\textbf{91} 141118 (2007).

\bibitem{Krause2008} M.~Krause, H.~Renner, and E.~Brinkmeyer,
  \textit{Optical isolation in silicon waveguides based on nonreciprocal Raman amplification},
  Electron.~Lett.~\textbf{44}, 691 (2008).

\bibitem{Poulton2010} C.~G.~Poulton, R.~Pant, A.~Byrnes, S.~Fan, M.~J.~Steel,
  and B.~J.~Eggleton,
  \textit{Design for broadband on-chip isolator using stimulated Brillouin scattering in dispersion-engineered chalcogenide waveguides},
  Opt.~Ex.~\textbf{20}, 21235 (2010).

\bibitem{Ramezani2010}
  H.~Ramezani, T.~Kottos, R.~El-Ganainy, and D.~N.~Christodoulides, {\it Unidirectional nonlinear PT-symmetric optical structures}, Phys. Rev. A {\bf 82}, 043803 (2010).

\bibitem{Miroschnichenko2010} A.~E.~Miroshnichenko, E.~Brasselet, and Y.~S.~Kivshar,
  \textit{Reversible optical nonreciprocity in periodic structures with liquid crystals},
  Appl.~Phys.~Lett.~\textbf{96}, 063302 (2010).

\bibitem{lepri2011prl}
  S.~Lepri and G.~Casati, \textit{Asymmetric wave propagation in nonlinear systems},
  Phys.~Rev.~Lett.~\textbf{106}, 164101 (2011).

\bibitem{anand2013nl} B.~Anand, R.~Podila, K.~Lingam, S.~R.~Krishnan, S.~Siva Sankara Sai, R.~Philip, and A.~M.~Rao, \textit{Optical diode action from axially asymmetric nonlinearity in an all-carbon solid-state device}, Nano Lett.~\textbf{13} 5771 (2013).

\bibitem{lepri2013pre}
  S.~Lepri and B.~A.~Malomed. \textit{Symmetry breaking and restoring wave transmission in diode-antidiode double chains}, Phys.~Rev.~E \textbf{87}, 042903 (2013).

\bibitem{xu2014prb}
  Xu, Yi, and Andrey E. Miroshnichenko, \textit{Reconfigurable nonreciprocity with a nonlinear Fano diode}, Physical Review B~\textbf{89}, 134306(2014).

\bibitem{li2014sp}
  N.~Li and J.~Ren, \textit{Non-Reciprocal Geometric Wave Diode by Engineering Asymmetric Shapes of Nonlinear Materials}, Sci.~Rep.~\textbf{4}, 6228 (2014).

\bibitem{peng2014nature} B.~Peng, S.~K.~{\"O}zdemir, F.~Lei, F.~Monifi,
  M.~Gianfreda, G.~L.~Long, S.~Fan, F.~Nori, C.~M.~Bender, and
  L.~Yang,
  \textit{Parity-time-symmetric whispering-gallery microcavities},
  Nat.~Phys.~\textbf{10}, 394 (2014).

\bibitem{chang2014nature} L.~Chang, X.~Jiang, S.~Hua, C.~Yang, J.~Wen,
  L.~Jiang, G.~Li, G.~Wang, and M.~Xiao,
  \textit{Parity-time symmetry and variable optical isolation in active-passive-coupled microresonators},
  Nat.~Phot.~\textbf{8}, 524 (2014).

\bibitem{Fan2015} Y.~Shi, Z.~Yu, and S.~Fan,
  \textit{Limitations of nonlinear optical isolators due to dynamic reciprocity},
  Nat.~Phot.~\textbf{9}, 388 (2015).

\bibitem{zhou2016oe} X.~Zhou and Y.~D.~Chong, \textit{PT symmetry breaking and nonlinear optical isolation in coupled microcavities}, Opt.~Ex.~\textbf{24}, 7, 6916 (2016).

\bibitem{jiang2016chip}
  X.~Jiang, C.~Yang, H.~Wu, S.~Hua, L.~Chang, Y.~Ding, Q.~Hua, and M.~Xiao,
  \textit{On-chip optical nonreciprocity using an active microcavity}, 
  Sci.~Rep.~\textbf{6}, 38972 (2016).


\bibitem{li2017ssh}
  C.~Li, \textit{et al.}, \textit{Unidirectional transmission in 1D nonlinear photonic crystal based on topological phase reversal by optical nonlinearity},
  AIP Advances \textbf{7}, 025203 (2017).

\bibitem{Sergey2016lpr}
S.~Sergey, A. ~Sukhorukov, J.~Huang, S.~Dmitriev, C.~Lee, and Y.~Kivshar, \textit{Nonlinear switching and solitons in PT-symmetric photonic systems},
  Laser and Phot.~Rev.~\textbf{10}, 177 (2016).

\bibitem{bernevigbook}
B.~A.~Bernevig and T.~L.~Hughes, \textit{Topological Insulators and Topological Superconductors} (Princeton University Press, 1984).

\bibitem{haldane2008prl}
  F.~D.~M.~Haldane and S.~Raghu, \textit{Possible realization of directional optical waveguides in photonic crystals with broken time-reversal symmetry}, Phys.~Rev.~Lett.~\textbf{100}, 013904 (2008).

\bibitem{haldane2008pra}
  S.~Raghu and F.~D.~M.~Haldane. \textit{Analogs of quantum-Hall-effect edge states in photonic crystals}, Phys.~Rev.~A \textbf{78} 033834 (2008).

\bibitem{wang2008prl}
  Z. Wang, Y. Chong, J.~D. Joannopoulos, and M.~Solja\u{c}i\'{c},
  \textit{Reflection-free one-way edge modes in a gyromagnetic photonic crystal}, Phys.~Rev.~Lett.~\textbf{100}, 013905 (2008).

\bibitem{wang2008nature}
  Z.~Wang, Y.~Chong, J.~D. Joannopoulos, and M.~Solja\u{c}i\'{c}, {\it Observation of unidirectional backscattering-immune topological electromagnetic states}, Nature {\bf 461}, 772 (2009).

\bibitem{hafezi2011nphy}
  M.~Hafezi, S.~Mittal, J.~Fan, A.~Migdall, and J.~M.~Taylor, {\it Imaging topological edge states in silicon photonics}, Nature Photon.~{\bf 7}, 1001 (2013).

\bibitem{fang2012nphoton}
  K.~Fang, Z.~Yu, and S.~Fan, \textit{Realizing effective magnetic field for photons by controlling the phase of dynamic modulation}, Nat.~Photon.~\textbf{6}, 782 (2012).

\bibitem{liang2013} G.~Q.~Liang and Y.~D.~Chong, \textit{Optical Resonator Analog of a Two-Dimensional Topological Insulator}, Phys.~Rev.~Lett.~\textbf{110}, 203904 (2013).

\bibitem{khanikaev2013nmat}
  A.~B. Khanikaev, S.~H. Mousavi, W.-K. Tse, M.~Kargarian, A.~H. MacDonald, and G.~Shvets, {\it Photonic topological insulators}, Nature Materials {\bf 12}, 223 (2013).

\bibitem{rechtsman2013nature}
  M.~C.~Rechtsman, J.~M.~Zeuner, Y.~Plotnik, Y.~Lumer, D.~Podolsky, F.~Dreisow, S.~Nolte, M.~Segev, and A.~Szameit, {\it Photonic Floquet topological insulators}, Nature {\bf 496}, 196 (2013).

\bibitem{hafezi2013} M.~Hafezi, S.~Mittal, J.~Fan, A.~Migdall, and J.~M.~Taylor, \textit{Imaging topological edge states in silicon photonics}, Nat.~Photonics {\bf 7}, 1001 (2013).

\bibitem{hafezi2014} S.~Mittal, J.~Fan, S.~Faez, A.~Migdall, J.~M.~Taylor, and M.~Hafezi, \textit{Topologically Robust Transport of Photons in a Synthetic Gauge Field}, Phys.~Rev.~Lett.~\textbf{113}, 087403 (2014).

\bibitem{lu2014nphoton}
  L.~Lu, J.~D. Joannopoulos, and M.~Solja\u{c}i\'{c}, {\it Topological photonics}, Nature Photon. {\bf 8}, 821 (2014).

\bibitem{Lumer2013}
  Y.~Lumer, Y.~Plotnik, M.~C.~Rechtsman, and M.~Segev, {\it Self-localized states in photonic topological insulators}, Phys.~Rev.~Lett.~{\bf 111}, 243905 (2013).

\bibitem{ablowitz2014}
M.~J.~Ablowitz, C.~W.~Curtis, and Y.-P.~Ma, {\it Linear and nonlinear traveling edge waves in optical honeycomb lattices}, Phys.~Rev.~A {\bf 90}, 023813 (2014).

\bibitem{alu2016} Y.~Hadad, A.~B.~Khanikaev and A.~Al\`u, \textit{Self-induced topological transitions and edge states supported by nonlinear staggered potentials}, Phys.~Rev.~B.~\textbf{93}, 155112 (2016).

\bibitem{daniel2016soliton}
  D.~Leykam and Y.~D.~Chong, \textit{Edge solitons in nonlinear-photonic topological insulators}, Phys.~Rev.~Lett.~\textbf{117}, 143901 (2016).

\bibitem{Makris2005} K.~G.~Makris, S.~Suntsov, D.~N.~Christodoulides, G.~I.~Stegeman, and A.~Hach\'e, \textit{Discrete surface solitons}, Opt.~Lett.~\textbf{30}, 2466 (2005).

\bibitem{Suntsov2006} S.~Suntsov, K.~G.~Makris, D.~N.~Christodoulides, G.~I.~Stegeman, A.~Hach\'e, R.~Morandotti, H.~Yang, G.~Salamo, and M.~Sorel, \textit{Observation of Discrete Surface Solitons}, Phys.~Rev.~Lett.~\textbf{96}, 063901 (2006).

\bibitem{Suntsov2007} S.~Suntsov, K.~G.~Makris, D.~N.~Christodoulides, G.~I.~Stegeman, R.~Morandotti, H.~Yang, G.~Salamo, and M.~Sorel, \textit{Power thresholds of families of discrete surface solitons}, Opt.~Lett.~\textbf{32}, 3098 (2007).

\bibitem{christodoulides1988} D.~N.~Christodoulides and R.~I.~Joseph, \textit{Discrete self-focusing in nonlinear arrays of coupled waveguides}, Opt.~Lett.~\textbf{13}, 794 (1988).

\bibitem{zeuner2015prl} J.~M.~Zeuner, M.~C.~Rechtsman, Y.~Plotnik, Y.~Lumer, S.~Nolte, M.~S.~Rudner, M.~Segev, and A.~Szameit, \textit{Observation of a topological transition in the bulk of a non-hermitian system}, Phys.~Rev.~Lett.~\textbf{115}, 040402 (2015).

\bibitem{weimann2016nmaterial} S.~Weimann, M.~Kremer, Y.~Plotnik, Y.~Lumer, S.~Nolte, K.~G.~Makris, M.~Segev, M.~C.~Rechtsman, and A.~Szameit, \textit{Topologically protected bound states in photonic parity-time-symmetric crystals}, Nat.~Mat.~\textbf{16}, 433 (2016).

\bibitem{cheng2015lp} Q.~Cheng, Y.~Pan, Q.~Wang, T.~Li, and S.~Zhu, \textit{Topologically protected interface mode in plasmonic waveguide arrays},
  Laser and Phot.~Rev.~\textbf{9}, 4, 392 (2015).

\bibitem{kitagawa2012ncom}Kitagawa, T., M. A. Broome, A. Fedrizzi, M. S. Rudner, E. Berg, I. Kassal, A. Aspuru-Guzik, E. Demler, and A. G. White, \textit{Observation of topologically protected bound states in photonic quantum walks}, Nat.~Comm.~\textbf{3}, 882 (2012).

\bibitem{leykam2016a}
D.~Leykam, M.~C.~Rechtsman, and Y.~D.~Chong, \textit{Anomalous Topological Phases and Unpaired Dirac Cones in Photonic Floquet Topological Insulators}, Phys.~Rev.~Lett.~\textbf{117}, 013902 (2016).

\bibitem{Haldane1988} F.~D.~M.~Haldane, {\it Model for a Quantum Hall Effect without Landau Levels: Condensed-Matter Realization of the ``Parity Anomaly''}, Phys.~Rev.~Lett.~{\bf 61}, 2015 (1988).

\bibitem{Melvin2006}
T.~R.~O. Melvin, A.~R. Champneys, P.~G. Kevrekidis, and J. Cuevas, {\it Radiationless traveling waves in saturable nonlinear Schr\"odinger lattices}, Phys. Rev. Lett. {\bf 97}, 124101 (2006).

\bibitem{Morandotti1999}
R.~Morandotti, U.~Peschel, J.~S.~Aitchison, H.~S.~Eisenberg, and Y.~Silberberg, {\it Dynamics of discrete solitons in optical waveguide arrays}, Phys. Rev. Lett. {\bf 83}, 2726 (1999).

\bibitem{liang2014} G.~Q.~Liang and Y.~D.~Chong, \textit{Optical Resonator Analog of a Photonic Topological Insulator: A Finite-Difference Time-Domain Study}, Int.~J.~Mod.~Phys.~B \textbf{28}, 1441007 (2014).

\bibitem{Pasek2014} M.~Pasek and Y.~D.~Chong, \textit{Network models of photonic Floquet topological insulators}, Phys.~Rev.~B \textbf{89}, 075113 (2014).

\bibitem{Hu2015} W.~Hu, J.~C.~Pillay, K.~Wu, M.~Pasek, P.~P.~Shum, and Y.~D.~Chong, \textit{Measurement of a Topological Edge Invariant in a Microwave Network}, Phys.~Rev.~X \textbf{5}, 011012 (2015).

\bibitem{Fei2016} F.~Gao, \textit{et al.}, \textit{Probing topological protection using a designer surface plasmon structure}, Nat.~Comm.~\textbf{7}, 11619 (2016).

\bibitem{nonlinear_lattices}
Y.~V. Kartashov, B.~A. Malomed, and L. Torner, {\it Solitons in nonlinear lattices}, Rev. Mod. Phys. {\bf 83}, 247 (2011).

\bibitem{boyd_book}
R.~W. Boyd, {\it Nonlinear Optics} (Academic, 2003).

\end{thebibliography}
\end{document}